# Orientation dependent work function of *in situ* annealed strontium titanate

L F Zagonel[1], M Bäurer[2], A Bailly[3], O Renault[3], M Hoffmann[2],
S-J Shih[4], D Cockayne[4], N Barrett[1]

[1] CEA DSM/IRAMIS/SPCSI, CEA Saclay, 91191 Gif sur Yvette, France
[2] Institut für Keramik im Maschinenbau, Universität Karlsruhe, D-76131 Karlsruhe, Germany
[3] CEA-LETI, MINATEC, 17 rue des Martyrs, 38054 Grenoble Cedex 9, France
[4] Department of Materials, University of Oxford, Oxford OX1 3PH, United Kingdom

Email: nick.barrett@cea.fr

Online at stacks.iop.org/JPhysCM/21/314013

**Abstract.** We have used energy filtered x-ray photoelectron emission microscopy (XPEEM) and synchrotron radiation to measure the grain orientation dependence of the work function of a sintered niobium doped strontium titanate ceramic. A significant spread in work function values is found. Grain orientation and surface reducing/oxidizing conditions are the main factors in determining the work function. Energy filtered XPEEM looks ideally suited for analysis of other technologically interesting polycrystalline samples.

**Keywords:** photoelectron spectroscopy, XPEEM, STO, work function, grain orientation.

## 1. Introduction

Oxide ceramics such as $SrTiO_3$ (STO) are widely used for passive devices in high frequency systems such as filters and antennas, and also in active devices such as tuneable rf filters and phase switchers. STO based compounds are also used in grain boundary layer capacitors, in which conductivity differences increase the device performance. The electronic structure of strontium titanate can be changed by the introduction of dopants, e.g. changing it from insulator to metallic conductor. [1] Recently, polycrystalline STO has been under discussion in the semiconductor industry as a candidate material for DRAM memories. [2] STO is also important as a model system for more complex oxides. The solid solution $STO$-$BaTiO_3$, for example, is used as a capacitor material. For applications in electronic devices, the contact between the ceramic and the electrode material plays an important role in device performance. [3] In particular, band alignment at the metal–oxide interface can determine the electrical properties of the device. [4]

The work function difference at an electrode–insulator contact [5] can give rise to the formation of electrical double layers, especially important in miniaturized devices where the ratio of surface to volume is very large. Due to work function differences at the interface, a positive or negative interface charge may be expected, thus directly affecting the performance of field effect devices and the oxidation of the metallic contact [6]. For example, in YBaCuO/STO, excess electrons are trapped on the YBCO side of the interface. [7] Shifts in the band alignment can lead to variations in the space-charge and thus in the interfacial electrostatic fields, which in turn influence the rate of diffusion of $O^{2-}$ ions and the electrode oxidation rate at the interface.

The work function is sensitive to the surface chemistry and STO surface stoichiometry is orientation dependent. Thus, the work function is expected to depend on the crystal orientation of the grains. Vlachos *et al.* showed that heating STO(100) above 1300 K reduces the surface via the creation of oxygen vacancies, provoking significant changes in the work function up to 0.82 eV. [8] Maus-Friedrichs et al observe surface defects after annealing in UHV at 1000K, giving rise to a density of states at 1-2 eV below the Fermi level with an intensity of 1% of the oxygen 2p valence band emission. [9]

In the literature, the work function of undoped STO (100) is given as 4.2 eV. [10] Similar values were reported for undoped and for 1at. % Nb doped material. [4] However, little data exists on the work function of sintered ceramics. [11] The work function may be measured using ultra-violet photoelectron spectroscopy (UPS) and a photon source such as He I. Applying a negative polarisation to the sample allows one to measure the spectral width between threshold and the Fermi level and thus deduce the work function, provided that the offset between the sample and spectrometer work functions is known. However, such UPS analysis will give, at best, an area averaged value for polycrystalline samples such as ceramics. In fact, different crystallographic grain orientations should give rise to a range of work function values. Furthermore, area averaged measurements cannot, by definition, distinguish contributions from the grain boundaries to the work function. In order to analyse single grain work functions it is necessary to combine spatial resolution of individual grains with the complete spectral information from UPS. Photoelectron emission microscopy with full energy filtering provides such a tool. If samples are sufficiently conducting, the work function is directly measured from the position of the photoemission threshold since the photoelectron energy E in PEEM is referenced with respect to the sample holder Fermi level $E_F$. Thus, $E-E_F=E_k+\Phi_{WF}$, where $E_k$ is the kinetic energy and $\Phi_{WF}$ the sample work function. [12] The technique therefore seems well adapted for a quantitative analysis of the grain orientation dependence of the work function in sintered niobium doped STO ceramics.

First, we describe the sample preparation and measurement methods. Then, the scanning electron microscopy and electron back-scattering diffraction results are presented on the morphology and crystal structure of the ceramic. Area averaged UPS valence band spectroscopy gives a first insight into the work function behaviour. Finally, a quantitative energy-filtered XPEEM analysis provides the orientation and chemical state dependent grain work function.

## 2. Experiment

High purity strontium carbonate (99.9%, Sigma Aldrich), titania (99.9%, Sigma Aldrich) and niobium oxide (99.9%, ChemPur) powders were used to prepare polycrystalline Nb doped STO powder. After milling and calcination, the powder was uniaxially pressed into discs of 15 mm diameter and 5 mm thickness in a steel die and subsequently cold isostatically pressed at 400 MPa. The discs were heated to 1700 K at 20 Kmin$^{-1}$ in a 95% $N_2$+5% $H_2$ atmosphere and then kept at temperature for 20 h. After sintering the sample was quenched at more than 200Kmin$^{-1}$. Further details on the synthesis procedure have been published elsewhere. [1] The sample diameter was reduced to 10.5 mm and then cut into discs of 1 mm thickness allowing the analysis of the central part. Finally, the sample was polished down to 0.25 µm diamond paste (no chemical polishing was used). Once in the vacuum chamber (base pressure of 2×10$^{-8}$ Pa), the sample was annealed for 3 hours at 1073 K. This annealing procedure has already been shown to yield a clean, ordered surface in single crystal test samples oriented in the major directions.

The X-ray photoelectron emission microscopy (XPEEM) was carried out using an energy-filtered microscope (NanoESCA, Omicron Nanotechnology GmbH). The instrument was installed at the ID08 line of the European Synchrotron Radiation Facility (ESRF) which delivers soft x-rays over the 400–1500 eV photon energy range. The threshold measurements were carried out using hv = 400 eV and an analyser pass energy of 100 eV, giving an overall energy resolution of 0.5 eV. All results were taken using a 35 µm field of view (FoV). The lateral resolution estimated from a fit of the 16-84% intensity rise across a grain boundary is 0.25 µm.

After XPEEM experiments the same sample area was observed using Scanning Electron Microscopy (SEM) and the grain orientation was determined by electron back-scattering diffraction (EBSD). The EBSD measurements were carried out at the Materials Department of the University of Oxford. The microstructure was studied using a scanning electron microscope (FEG LEO 1525) at the CEA Saclay. High quality Kikuchi patterns were easily obtained without any further sample preparation. Reference marks placed on the surface allowed individual grains to be identified in the XPEEM experiments and in the EBSD observations, in order to determine the grain orientation for each work function measurement.

The area averaged electronic structure was studied with standard laboratory ultra-violet photoelectron spectroscopy (UPS) using He I light (hν = 21.2 eV). The probed area was of about 1 mm$^2$, thus averaging the grain orientation distribution. The electron energy analyser was set to a resolution of 50 meV and operated at normal detection angle. The photoemission threshold region was measured using -6 V sample polarization with respect to the analyzer.

## 3. Results and discussion

### 3.1 Morphology

Figure 1(a) shows the microstructure of the sample as determined by SEM. This is a backscattered electron image (secondary electron images showed no contrast), taken at an electron acceleration voltage of 15kV. The sintering in hydrogen atmosphere considerably inhibits the grain growth. As a consequence, the grain size is rather small compared to the average size observed on samples prepared in similar conditions but in air atmosphere. [1] A higher porosity is also observed again with respect to samples prepared in air. Figure 1(b) shows the EBSD results from the same region of the sintered STO ceramic, the measured grain orientations follow the colour code in the inset.

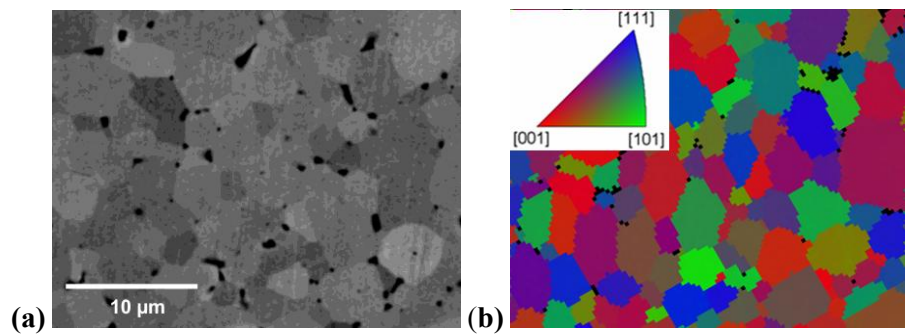

**Figure 1.** (a). SEM micrograph of the Nb doped STO sample after sintering in H$_2$ atmosphere for 20 hours. The field of view used is the same as in the XPEEM experiments. Primary energy was 15 keV (b) EBSD image together with the grain orientation colour coding.

### 3.2 Average valence band structure

The valence band of the STO ceramic is dominated by the O 2p band at a binding energy of ~ 7 eV, as shown in figure 2. Auger electron spectroscopy shows that annealing eliminates the surface carbon contamination, however, annealing can also generate oxygen vacancies. Comparison of annealing in oxygen and in ultra high vacuum (UHV) shows clearly the reducing influence of the latter. [13, 14] Features at low binding energies are present in the bulk gap due to oxygen vacancies induced by UHV annealing. [15] Thus, we expect that those grain orientations with oxygen rich surface terminations will be likely to show more important variations with respect to the nominal work function for stoichiometric surfaces. In a careful UPS study of single crystal STO (100) surfaces prepared under reducing and oxidising conditions, Aiura *et al* [16] observe a sharp metallic peak at the Fermi cut-off in the Γ-Δ-X direction with a resonant behaviour. This metallic peak is well off resonance for He I radiation, furthermore it is unlikely that one of the STO grains in our ceramic sample is [001] oriented with the same angular precision as that employed in UPS. However, we do observe a non-zero density of states in the bulk band gap, with similar intensity to that reported by Kido *et al*. [15] and Aiura *et al*. [16]

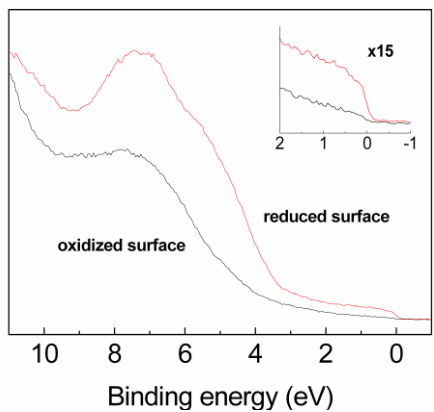

**Figure 2.** Area averaged UPS spectra of the STO ceramic as a function of surface treatment. In black, the spectrum after annealing at 1073 K for 30 minutes in $2\times10^{-8}$ mbar $O_2$ (36 Langmuir). In red, the spectrum after annealing for 90 minutes in UHV. The inset shows a zoom on the region around the Fermi level. The density of states in the bulk band gap between the oxygen 2p and the Fermi level of the oxidized surface points to residual carbon contamination on the STO grains or due to the porosity, whereas the much higher density of states and the clear Fermi level cut-off of the reduced surface suggests the presence of oxygen vacancies at the surface of the grains.

### 3.3 Work-Function

The work function was determined for 62 grains using XPEEM just after an in situ UHV annealing in UHV at 1073 K, i.e. in the same conditions as for the "reduced surface" of Fig. 2. For this measurement a series of images was taken by sweeping $E-E_F$ through the photoemission threshold region from 3 to 10 eV in 0.05 eV steps. One example of such an image is show in Figure 3(a), recorded at the very beginning of the photoemission threshold with $E-E_F$ = 3.85 eV. Darker grains indicate higher values of work-function while brighter grains have lower values of work-function. The clear presence of an energy distribution of photoelectron emission around the generally accepted work function values of 4.2 eV is due to the band pass window of the energy analysis (0.5 eV). The grain spectra are extracted from the image series by defining an area of interest within each grain. After correction for the Schottky effect due to the high extractor field, ΔE=98 meV for 12 kV, [17] the position of the photoemission threshold corresponds to the work function. This is a calculated correction which assumes that the Nb doping makes the STO sufficiently conducting to provide a metal like image potential; the actual values may be a few tens of milli-electronvolts higher. Secondly, one would expect rumpling for the grains with a mixed oxygen-cation termination, [18] leading to an outwards relaxation of the oxygen atoms, hence further compensating the Schottky effect. On the other hand, purely cation terminated surfaces should enhance the work function reduction. In both cases, it seems reasonable to treat the reported work function values as lower limits.

The opening angle α of the energy analyser gives rise to a parabolic energy dispersion or isochromaticity, in the vertical plane as a function of position in a given FoV as shown in figure 3(b). The isochromaticity correction to the $E-E_F$ scale in the image series is obtained by integrating the image intensities over the FoV for constant y values and includes a vertical offset due to the instrumental alignment, apparent in figure 3(a) where the upper part of the FoV is brighter than the lower part. Figure 3(c) shows an example of the fit to the corrected experimental data using an error function to simulate the rising edge of the photoemission threshold. The uncertainty in the error function fits was ~ 0.01 eV. The average work function extracted from all 62 grains was (4.26 ± 0.06 eV). Thus the spread in work function values as a function of grain orientation is clearly significant with respect to the uncertainty in the fits to the individual threshold spectra.

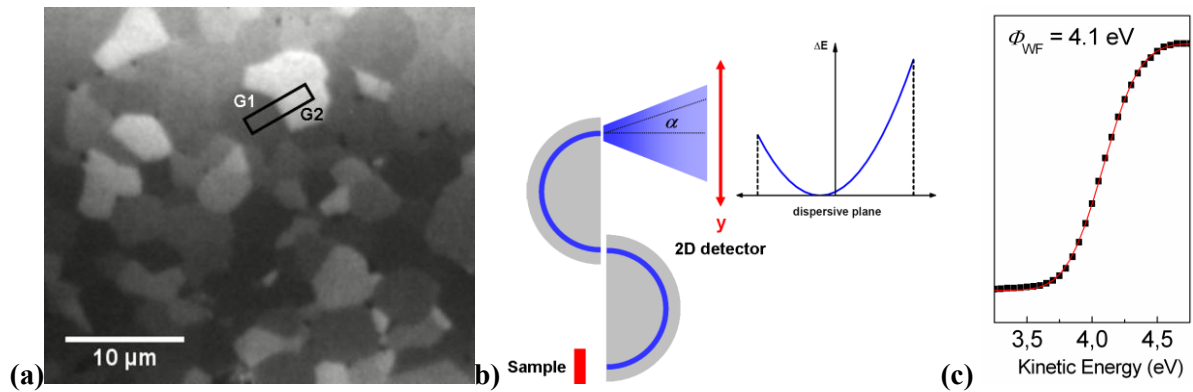

**Figure 3.** (a) Raw data image acquired at the photoemission threshold (E-$E_F$ = 3.85 eV) with a field of view 35 μm. Due to the energy resolution of 0.5 eV some grains already emit photoelectrons. The isochromaticity in the dispersive plane of the analyser is apparent in the upper, lighter, part of the figure. Grains G1 and G2 are used for the grain boundary analysis below. (b) Schematic showing the dispersive plane in the NanoESCA giving rise to a parabolic correction to the FoV energy dispersion applied to all images before spectral extraction. (c) Typical threshold spectrum extracted from a 150 nm × 150 nm area of interest and the best least squares fit using an error function.

The EBSD data of figure 1(b) can be represented in terms of polar (θ) and azimuthal (φ) angles. No correlation between the extracted work function values from XPEEM and φ was found. However, there is a high linear correlation (0.85) between θ and the work function, as can be seen from figure 4(a). In particular, the work function increases as the grain orientation passes successively through the three principal crystallographic directions <100>, <110> and <111>. Figure 4(b) is a histogram showing the number of grains observed in each 0.05 eV work function interval. As indicated, the average value is 4.26 eV with a standard deviation of 0.06 eV while grains were observed with work functions ranging from 4.10 eV up to 4.35 eV. The median work function is observed in the 4.25-4.30 eV interval and the entire histogram is skewed, with a longer tail at low work function values. The work function obtained from the area averaged UPS after annealing in UHV was 4.34 eV. The small difference with the mean XPEEM measured value of 4.26 eV may be due to slightly different reducing conditions, or possibly to the approximation used to correct for the Schottky effect on the measured work function

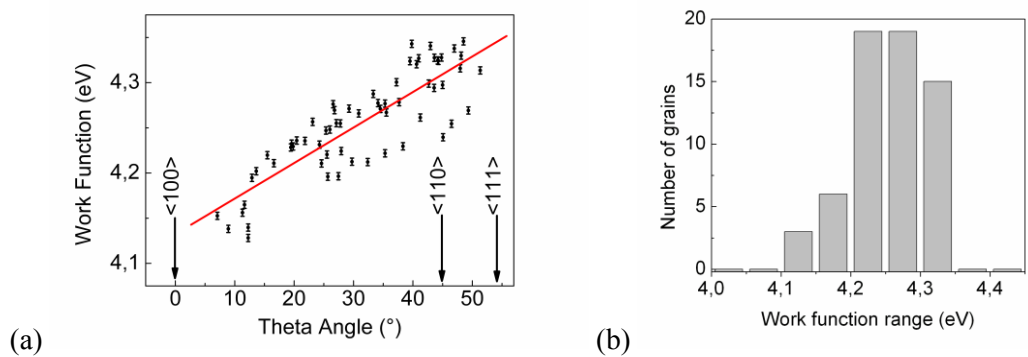

**Figure 4.** (a) Experimentally determined work function as a function of the θ angle as defined by electron back-scattering diffraction. The error bars correspond to a standard deviation of 0.06 eV for the error function fit to each photoemission threshold spectrum. The red line is the best linear correlation between work function and θ, with an R-value of 0.85. (b) Histogram of the distribution of work function values for the 62 analysed grains. The mean value is 4.26 eV.

The linear increase of the STO grains work function as function of the orientation angle should be treated as an overall consequence of surface layer structural and chemical changes which determine the surface electronic properties. Each grain surface will have its own particular termination (possibly with a reconstruction) and stoichiometry as observed for different single crystals. [6,19-21] In an XPEEM study of the surface termination chemistry of the *same* ceramic, we concluded that the [100] oriented grains were preferentially terminated by SrO, whilst the polar [110] and [111] orientations seamed to have, respectively, $O_2^{4-}$ and $Ti^{4+}$ termination layers. [22] These results are also an overall estimate showing different chemical properties of different orientations, although we cannot exclude a contribution from surface reconstruction.

The UPS valence band study indicates slight surface oxygen depletion. This surface reduction will affect principally the [100] and [110] orientations, providing an explanation for the low energy tail in the work function histogram.

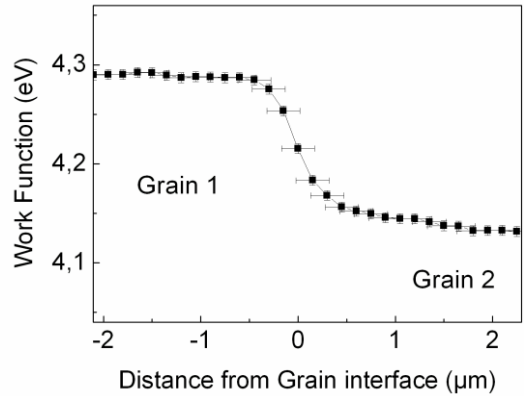

**Figure 5.** Work function obtained from least square error function fit to the extracted threshold spectra from 150 nm wide regions at 170 nm intervals across the grain boundary G1/G2 shown in figure 3(a). The maximum error bar in the work function value is ±0.01 eV

Figure 5 shows the work function obtained from energy filtered photoemission threshold spectra extracted across a grain boundary between grains with θ orientation within ~ 10° of [100] and [101]. The transition in work function values between neighbouring grains is smooth, making the existence of a distinct phase within the grain boundary unlikely. For the area averaged UPS work function measurement such a contribution would in any case be negligible since the grain boundaries account for less than 1% of the analyzed area. With the spatial resolution of XPEEM the GB contribution could attain 5-10% of the extracted signal. In the extreme case of two distinguishable work functions one would expect structure in the extracted threshold spectra, as we have already observed on single gold-covered Si nanowires [12] or a significant broadening of the standard deviation in the error function fit to the rise in the photoemission signal. This is clearly not the case; thus we can exclude the grain boundaries as the source of the difference between the work function measured by XPEEM and that measured a posteriori by area averaged UPS. As in the case of single crystal STO studies, [4, 10] it is the oxidizing/reducing conditions during the surface preparation and the chemistry of the terminating layer of each grain which play the dominant roles. The transition region between the two grains is 0.30-0.35 μm wide, slightly larger than the spatial resolution (0.25 μm). Thus, we cannot exclude the existence of a concentration gradient in charged defects due to a space charge region around the grain boundary. However, visualizing the space charge is outwith the spatial resolution used in these experiments. Nevertheless, there is a wealth of microscopic information well beyond a classical area averaged UPS analysis, underlining the importance of being able to carry out high resolution spatially resolved electron spectroscopy on a significant field of view of the sintered ceramic.

**Conclusions**

We have used fully energy filtered X-ray photoelectron emission microscopy and synchrotron radiation to measure the gain orientation dependent work function $\Phi_{WF}$ of a niobium doped STO sintered ceramic. Photoemission threshold spectra are extracted from 62 distinct grains. The mean work function value is 4.26 eV with a standard deviation of 0.06 eV. The grain orientation is determined by electron back-scattered diffraction and correlated with the XPEEM work functions. A good linear correlation is obtained between $\Phi_{WF}$ and the polar EBSD angle θ. In particular, the work functions determined for the three principal crystal faces [100], [110] and [111] are 4.13, 4.32 and 4.34 eV, respectively. The oxygen rich SrO terminated [100] face is particular sensitive to the surface preparation conditions used in-situ. In our work, the UHV annealing has produced surface defects, probably oxygen vacancies, evidenced by the existence of a non-zero density of states in the bulk band gap, as measured by a posteriori UPS. Further confirmation of the reduced nature of the XPEEM analyzed surface is obtained from the area averaged work function of 4.48 eV as determined by UPS on the ceramic after oxidation. Energy filtered XPEEM thus appears to be a powerful tool for quantitative analysis of spatially varying work functions in technologically interesting polycrystalline materials.

It would be particularly interesting to combine in a single experiment spatially resolved work function and chemical state analysis with either grain sensitive photoelectron diffraction an/or low energy electron microscopy (LEEM), in order to correlate the finer aspects of the links between surface electronic and atomic structure.

### Acknowledgments

This work was supported by the European commission under contract Nr. NMP3-CT-2005-013862 (INCEMS) and by the French national research agency project 05-NANO-065 XPEEM. The authors thank Patrick Bonnaillie for his skilled operation of the SEM and the ESRF-ID08 beam staff for their valuable technical assistance. We also thank B. Krömker for useful discussions on the intrinsic energy dispersion in the PEEM field of view.